\documentclass[prx,aps,superscriptaddress,floatfix,tightenlines,showpacs,twocolumn]{revtex4-2}
\usepackage{graphicx}
\usepackage{dcolumn}   
\usepackage{bm}        
\usepackage{amssymb}   
\usepackage{amsmath}
\usepackage{url}
\usepackage{layout}
\usepackage{color}
\usepackage{epsfig}
\usepackage{graphicx}
\usepackage{mathrsfs}\usepackage{bm}\usepackage{appendix}\usepackage{color}

\usepackage[thinlines]{easytable}
\usepackage{makecell}
\usepackage{tikz,pgf}
\usepackage{booktabs}
\usepackage{float}
\usepackage{hyperref}
\hypersetup{colorlinks,allcolors=blue}

\def\be{\begin{equation}}       \def\ee{\end{equation}}
\def\bea{\begin{eqnarray}}      \def\eea{\end{eqnarray}}
\def\bp{\begin{pmatrix}} \def\ep{\end{pmatrix}}
\def\beaa{\begin{equation}\begin{aligned}}
		\def\eeaa{\end{aligned}\end{equation}}

\begin{document}

\title{Interplay of two $E_g$ orbitals in Superconducting La$_3$Ni$_2$O$_7$ Under Pressure}

\author{Chen Lu}
\thanks{These two authors contributed equally to this work.}
\affiliation{New Cornerstone Science Laboratory, Department of Physics, School of Science, Westlake University, Hangzhou 310024, Zhejiang, China}
\author{Zhiming Pan}
\thanks{These two authors contributed equally to this work.}
\affiliation{Institute for Theoretical Sciences, Westlake University, Hangzhou 310024, Zhejiang, China}
\affiliation{New Cornerstone Science Laboratory, Department of Physics, School of Science, Westlake University, Hangzhou 310024, Zhejiang, China}
\author{Fan Yang}
\email{yangfan\_blg@bit.edu.cn}
\affiliation{School of Physics, Beijing Institute of Technology, Beijing 100081, China}
\author{Congjun Wu}
\email{wucongjun@westlake.edu.cn}
\affiliation{New Cornerstone Science Laboratory, Department of Physics, School of Science, Westlake University, Hangzhou 310024, Zhejiang, China}
\affiliation{Institute for Theoretical Sciences, Westlake University, Hangzhou 310024, Zhejiang, China}
\affiliation{Key Laboratory for Quantum Materials of Zhejiang Province, School of Science, Westlake University, Hangzhou 310024, Zhejiang, China}
\affiliation{Institute of Natural Sciences, Westlake Institute for Advanced Study, Hangzhou 310024, Zhejiang, China}

\begin{abstract}
The discovery of high-$T_c$ superconductivity (SC) in La$_3$Ni$_2$O$_7$ (LNO) has aroused a great deal of interests. 
Previously, it was proposed that 
the Ni-$3d_{z^2}$ orbital is crucial to realize the high-$T_c$ SC in LNO:
The preformed Cooper pairs therein
acquire coherence via hybridization with the $3d_{x^2-y^2}$ orbital to form the SC. 
However, we held a different viewpoint that 
the interlayer pairing $s$-wave SC is induced by the $3d_{x^2-y^2}$ orbital, driven by the strong interlayer superexchange interaction. 
To include effects from both $E_g$-orbitals , we establish a two-orbital bilayer $t$-$J$ model.
Our calculations reveal that due to the no-double-occupancy constraint, the $3d_{x^2-y^2}$ band and the  $3d_{z^2}$ bonding band are flattened by a factor of about 2 and 10, respectively, which is consistent with recent angle-resolved-photo-emission-spectroscopy measurements. 
Consequently, a high temperature SC can be hardly
induced in the $3d_{z^2}$-orbital due to the difficulty to develop phase coherence.
However, it can be easily achieved by the $3d_{x^2-y^2}$ orbital under realistic interaction strength. 
With electron doping, the 
$3d_{z^2}$-band gradually dives  below the Fermi level, but $T_c$ continues to enhance, suggesting that it 
is not necessary for the high-$T_c$ SC in LNO. 
With hole doping, $T_c$ initially drops and then rises, accompanied by the crossover from the
BCS to BEC-type superconducting transitions. 
\end{abstract}\maketitle


\section{Introduction}
The recent discovery of high temperature ($T_c\approx 80$K) superconductivity (SC) in La$_3$Ni$_2$O$_7$ (LNO) under pressures over $14$GPa ~\cite{Wang2023LNO} has aroused intensive interests, both experimental ~\cite{WenHH2023,Wang2023LNOb,YuanHQ2023LNO,yang2023arpes,zhang2023pressure,wang2023LNOpoly,wang2023la2prnio7,wang2023structure,zhou2023evidence,cui2023strain} and theoretical ~\cite{YaoDX2023,Dagotto2023,WangQH2023,lechermann2023,Kuroki2023,HuJP2023,ZhangGM2023DMRG,Werner2023,shilenko2023correlated,WuWei2023charge,cao2023flat,chen2023critical,YangF2023,lu2023bilayertJ,zhang2023structural,oh2023type2,liao2023electron,qu2023bilayer,Yi_Feng2023,jiang2023high,zhang2023trends,huang2023impurity,qin2023high,tian2023correlation,lu2023sc,jiang2023pressure,kitamine2023,luo2023high,zhang2023strong,pan2023rno,sakakibara2023La4Ni3O10,lange2023mixedtj,geisler2023structural,yang2023strong,rhodes2023structural,lange2023feshbach,labollita2023ele,kumar2023softening,kaneko2023pair,ryee2023critical,zhang2023la3ni2o6,Grusdt2023lno03349,chen2023iPEPS,liu2023dxy,ouyang2023hund,qu2023roles,sui2023rno,zheng2023twoorbital,kakoi2023pair,heier2023competing}. 
LNO and its possible generalizations R$_3$Ni$_2$O$_7$ (R=Rare earth element) \cite{zhang2023trends,pan2023rno,geisler2023structural} serve as a newly platform for high-$T_c$ SC other than the cuprates \cite{bednorz1986LBCO}. 
Although LNO hosts the conducting NiO$_2$ plane isostructural with the CuO$_2$ plane in cuprates, two significant differences exist between these two families in the aspect of crystal and electronic structures.

Firstly, while cuprate material is either a monolayer, or, multilayer system with weak interlayer couplings \cite{ubbens1994,kuboki1995,maly1996,nazarenko1996bilayer3d,medhi2009,eder1995,vojta1999,zhao2005,zegrodnik2017,matx2022}, the LNO possesses two NiO$_2$ layers within a unit cell hybridized via the Ni-O-Ni bonding assisted by the O-$2p$ orbital of the apical oxygen atoms. 
Note that although LNO is an existing material studied previously \cite{taniguchi1995transport,seo1996electronic,kobayashi1996transport,greenblatt1997ruddlesden,greenblatt1997electronic,ling2000neutron,wu2001magnetic,fukamachi2001nmr,voronin2001neutron,Bannikov2006,hosoya2008pressure,pardo2011dft,nakata2017finite,mochizuki2018strain,li2020epitaxial,song2020structure,barone2021improved,Wang2022LNO}, its properties undergo a fundamental change due to the structure transition from the orthorhombic structure with space group $Amam$ \cite{Wang2022LNO} to $Fmmm$ under pressure.
The Ni-O-Ni bonding angle changes from 168$^\text{o}$ to 180$^\text{o}$ enhancing the interlayer hybridization, after which high-T$_c$ SC emerges. 
This fact implies the crucial importance of the 
interlayer coupling for the emergence of high-T$_c$ SC in the pressurized LNO. 
Studies on bilayer Hubbard modes exist in literature.
In particular, an extended version of bilayer Hubbard 
model shows high-T$_c$ superconductivity arising from doping antiferrmagnetic Mott insulators by sign-problem free quantum Monte Carlo (QMC) simulations,
nevertheless, a strong inter-layer antiferromagnetic superexchange is required \cite{matx2022}.

Secondly, while cuprate can essentially be treated as a single orbital system, the LNO with Ni$^{2.5+}$ electron configuration simultaneously hosts two low-energy $E_g$ orbitals, i.e., the nearly 
half-filled $3d_{z^2}$ orbital and the nearly quarter-filled $3d_{x^2-y^2}$ orbital~\cite{pardo2011dft,nakata2017finite,Wang2023LNO,YaoDX2023,Dagotto2023,yang2023arpes}. 
Both orbitals are important in LNO: while the $3d_{z^2}$ orbital connects two NiO$_2$ layers via the Ni-O-Ni bonding, the $3d_{x^2-y^2}$ orbital is responsible for the intralayer transport.
At a glance, knowledge from cuprates \cite{kotliar1988,keimer2015highTc,proust2019highTc} seems to suggest that neither the half-filling nor the quarter-filling can provide an ideal platform to realize high-$T_c$ SC. 
However, the situations might be dramatically changed when the interplay between the two $E_g$ orbitals and the bilayer structure are taken into account. 
 On the one hand, there exists the nearest-neighbor (NN) bond hybridization between the two $E_g$ orbitals, which dictates that the concentrate of charge carrier is transformed between two orbitals. 
On the other hand, Hund's rule coupling polarizes electrons in two orbitals on the same site, 
which induces considerable inter-layer antiferromagnetic exchange between $3d_{x^2-y^2}$ orbitals intermediated with the $d_{z^2}$ orbitals ~\cite{lu2023bilayertJ}.

There are often two different viewpoints on the strong-coupling
analysis to the high-$T_c$ SC in the pressurized LNO: Which
orbital plays the primary role? 
One viewpoint is that the $3d_{z^2}$ orbital dominates \cite{cao2023flat,Yi_Feng2023,qin2023high}.
Since its orbital orientation is along the $z$-axis, the interlayer antiferromagnetic (AFM) superexchange between $3d_{z^2}$-orbitals is large.
Two electrons in the $3d_{z^2}$-orbital along the rung in a unit cell already pair below a high temperature scale $T^*$. 
However, these pairs still need to hop in the NiO$_2$ planes to develop phase coherence.
Two difficulties arises in this regard: one is the low hole density of the $3d_{z^2}$-orbital, and the other is the weak intralayer hopping integral between the adjacent $3d_{z^2}$-orbitals.
It was proposed~\cite{Wang2023LNO, ZhangGM2023DMRG} that pressure
could enhance the hybridization between two $E_g$ orbitals and lift up the $\sigma$-bonding band consisting of the $3d_{z^2}$ orbital to cross the Fermi level, {\it i.e.} to metallize, which could settle both difficulties, paving way for the $3d_{z^2}$-orbital to take 
the dominant role in the SC.

In contrast, the other viewpoint emphases on the $3d_{x^2-y^2}$ orbital  \cite{lu2023bilayertJ,qu2023bilayer,oh2023type2}. 
The key observation is the Hund's rule coupling between two $E_g$ orbitals on the same Ni site.
Integrating out the nearly localized spin moment of the $3d_{z^2}$
orbital under the strong Hund's rule coupling, an effective strong 
inter-layer superexchange between the $3d_{x^2-y^2}$ orbitals is generated. 
Therefore, an effective bilayer single-$3d_{x^2-y^2}$ orbital $t$-$J_{\parallel}$-$J_{\perp}$ model is obtained \cite{lu2023bilayertJ}. 
In addition to the conventional intra-layer superexchange $J_{\parallel}$, two layers are coupled by the strong inter-layer superexchange $J_{\perp}$.
$J_{\perp}$ favors an inter-layer $s$-pairing SC and enhances 
the $T_c$ dramatically\cite{lu2023bilayertJ,matx2022}. 
In this point of view, the role of pressure lies in that it
changes the lattice structure and therefore
significantly enhances the interlayer superexchange.

In this work, in order to study both $E_g$ orbitals on equal footing, we establish an effective two-orbital bilayer $t$-$J$ model to study their interplay, treated by the slave-boson mean field (SBMF) analysis \cite{kotliar1988,lee2006htsc}. 
The first result is that due to the no-double-occupancy constraint, the $3d_{x^2-y^2}$ band and the $\sigma$-bonding $3d_{z^2}$ band are flattened by a factor of about 2 and 10 respectively, which is well consistent with the recent angular resolved photo emission spectroscopy 
(ARPES) measurement \cite{yang2023arpes}. 
The second result is that the interlayer $s$-wave SC in LNO is induced by the pairing of the $3d_{x^2-y^2}$ orbital. 
For the $3d_{z^2}$ orbital, as the no-double-occupancy constraint strongly suppresses its hybridization with the $3d_{x^2-y^2}$ orbital, it can hardly acquire sufficient coherence to account for the high $T_c\approx 80$ K comparable with experiment, which however is tenable for the $3d_{x^2-y^2}$ orbital under realistic interaction parameters. 
The final result is the doping {\it versus} temperature phase diagram. With the electron doping the $T_c$ monotonically increases although the $\sigma$-bonding band gradually dives below the Fermi level, suggesting that the metallization of the bonding band is not necessary for the high-$T_c$ SC in LNO. 
With the hole doping the $T_c$ initially decreases and then increases, and the superconducting transition is initially the BCS type and then
crossover to the BEC type. 
Above $T_c$, there exists a pseudo-gap phase characterized by the pairing of the $3d_{z^2}$ spinons, which merges into the 
superconducting phase after some critical electron doping level. 
Such an interesting phase diagram with coexisting $T_c$ and $T^*$ appeals for experimental verification in the near future.

The rest part of this article is organized as follow. 
In Sec.~\ref{sec:effective_model}, the effective bilayer two-orbital model for LNO is established.
In Sec.~\ref{sec:SBMF_Band}, we introduce the SBMF approach to treat the two-orbital model, and the band renormalization effect 
is further revealed by the electron correlation.
In Sec.~\ref{sec:SC}, the SC state of the two-orbital model is studied. We shall first study the SC induced by the two orbitals separately and then clarify the obtained SC in the combined system. 
In Sec.~\ref{sec:Doping_Effect}, the effect of possible chemical doping in LNO is explored and the phase diagram including the pseudo gap phase
is obtained.
The conclusion is presented in Sec.~\ref{sec:Conclusion}.

\section{Effective bilayer two-orbital model}
\label{sec:effective_model}

The electronic properties of the LNO mateiral are determined by 
the $E_g$ orbitals of two Ni$^{2.5+}$ cations along the rung 
connecting the NiO$_2$ bilayer planes, whose energy bands 
are close to the Fermi energy.
Below the bilayer two-orbital model are constructed to describe 
the band structure and super-exchange physics . 

\subsection{The bilayer two-orbital Hubbard Hamiltonian}
The different orientations of two $E_g$ orbitals result in the strong inter-layer hopping $t^{z}_{\perp}$ between two $3d_{z^2}$ orbitals
along the rung, and also the relatively strong intra-layer
hopping $t^{x}_{\parallel}$ between the $3d_{x^2-y^2}$ orbitals.
The $3d_{z^2}$ and $3d_{x^2-y^2}$ orbitals within the intra-layer nearest
neighbor (NN) bond further hybridize,
but those along the rung do not. 
The corresponding band structure is formulated in the following tight-binding Hamiltonian,
\begin{widetext}
\begin{equation}
\begin{aligned}
H_0&=-t^{x}_{\parallel} \sum_{\langle i,j\rangle \alpha\sigma} \big(d_{x^2\alpha\sigma}^{\dagger} (i) d_{x^2\alpha\sigma} (j) +\text{h.c.}\big) 
+\Delta_g \sum_{i\alpha\sigma} d_{x^2\alpha\sigma}^{\dagger} 
(i)
d_{x^2\alpha\sigma} (i)
\\ 
&-t^{z}_{\perp} \sum_{i\sigma} \big(d_{z^21\sigma}^{\dagger} (i) d_{z^22\sigma} (i)
+\text{h.c.}\big) 
-t^{z}_{\parallel} \sum_{\langle i,j\rangle\alpha\sigma} \big(d_{z^2\alpha\sigma}^{\dagger} (i)
d_{z^2 \alpha\sigma} (j)
+\text{h.c.}\big)  -
\sum_{\langle i,j\rangle \alpha\sigma} t_{\parallel,j-i}^{xz} \big(d_{x^2\alpha\sigma}^{\dagger} (i)
d_{z^2 \alpha\sigma} (j) +\text{h.c.}\big)
\label{eq:ham0}
\end{aligned}
\end{equation}
\end{widetext}
where $i$ is the lattice site index; $\alpha$ 
represents the layer index; 
$\sigma$ is the spin component index.
$d_{a \alpha\sigma}^{\dagger}
(i)$ represents the electron creation operator for the $3d_{x^2-y^2}$ orbital ($a=x^2$) or 
the $3d_{z^2}$ orbital ($a=z^2$) with $a$ referring to the orbital index.
The $3d_{z^2}$ orbital has a smaller on-site energy as described by the $\Delta_g$ term.
The intra-layer hybridization $t_{\parallel,j-i}^{xz}$ between  $3d_{x^2-y^2}$ and $3d_{z^2}$ orbitals
on neighboring sites is comparable to hopping strengthes $t^{x}_{\parallel}$ and $t^{z}_{\perp}$.
The NN hybridization takes opposite signs along the $x$ and $y$ direction, respectively, 
i.e.,  $t^{xz}_{\parallel,x}=-t^{xz}_{\parallel,y}\equiv t^{xz}_{\parallel}$, due 
to the different symmetries of orbital wavefunctions.
A finite but small intra-layer hopping $t^{z}_{\parallel}$ for the $3d_{z^2}$ orbital is also added into Eq.~(\ref{eq:ham0}). 

The multi-orbital Hubbard interaction reads,
\begin{equation}
\begin{aligned}
&H_{1}= U\sum_{ia} n_{a\alpha\uparrow}(i)
n_{a\alpha\downarrow}(i) 
+V \sum_{i\alpha} n_{z^2\alpha}(i) n_{x^2\alpha}(i)
\\
&+P \sum_{i\alpha} \left( d^\dagger_{x^2\alpha\uparrow}(i) d^\dagger_{x^2\alpha \downarrow}(i)
d_{z^2\alpha \downarrow}(i) d_{z^2 \alpha 
\uparrow}(i) +\text{h.c}\right) \\
&-J_{H} \sum_{i\alpha} \bm{S}_{z^2\alpha}(i) \cdot \bm{S}_{x^2\alpha}(i),
\end{aligned}
\end{equation}
where the $U$- and $V$-terms describe the intra-orbital and inter-orbital repulsions, respectively; 
the $P$-term describes the singlet pair hopping between two orbitals;
the $J_H$-term is the Hund's rule coupling aligning
electron spins in two onsite $E_g$ orbitals;
$n_{s\alpha\sigma}(i)=d_{s\alpha\sigma}^{\dagger}(i)d_{s\alpha\sigma}(i)$ represents the particle number operator for an electron with spin-$\sigma$ in orbital $s$ with
$s=x^2$ or $z^2$;
$n_{s\alpha}(i)=n_{s\alpha\uparrow}(i)+n_{s\alpha\downarrow}(i)$;
$\bm{S}_{s\alpha}(i)=\frac{1}{2}d_{s\alpha}^{\dagger}(i)[\bm{\sigma}]d_{s\alpha}(i)$ is the spin operator for the 
electron in orbital $s$, where
Pauli matrices $\bm{\sigma}=(\sigma_x,\sigma_y,\sigma_z)$.

\subsection{The bilayer $t$-$J$ model}
In the strong coupling limit, the Hubbard-$U$ interaction generates an effective inter-layer AFM spin superexchange $J_{\perp}$ for the $3d_{z^2}$ orbitals and an intra-layer one $J_{\parallel}$ for the $3d_{x^2-y^2}$ orbitals, respectively.
Two electrons in the same site but different $E_g$ orbitals are aligned in their spin direction due to the strong Hund's coupling.
Consequently, the inter-layer superexchange $J_{\perp}$ of $3d_{z^2}$ orbital is transmitted to the $3d_{x^2-y^2}$ orbital \cite{lu2023bilayertJ}.
Please note that this interlayer AFM coupling does not arise from the virtual hopping due to
the inter-layer tunneling between two $3d_{x^2-y^2}$ orbitals, which is negligible. 

\begin{figure}[t]
\includegraphics[width=0.7\linewidth]{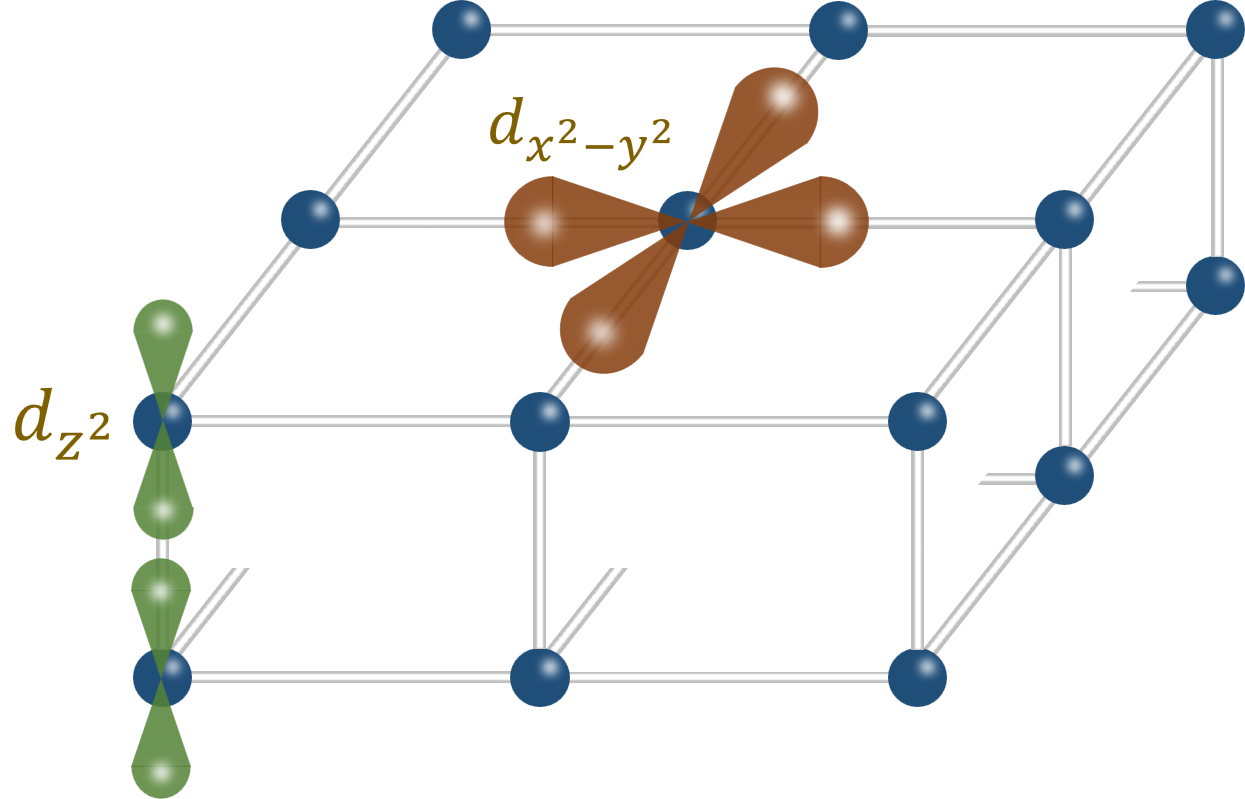}
\caption{Schematic figure of the bilayer $E_g$ orbital model.}
\label{fig:EgLattice}
\end{figure}

We arrive at an effective bilayer $t$-$J$ two-orbital model as depicted in Fig.~\ref{fig:EgLattice}:
The band part, or, the $t$-term, looks the same as Eq. (\ref{eq:ham0}) but under the constraint of no double-occupancy in each orbital.
The interaction part reads
\begin{equation}
\begin{aligned}
H_{J,x^2}&=J_{\parallel} \sum_{\langle i,j\rangle \alpha} \bm{S}_{x^2\alpha}(i) \cdot \bm{S}_{x^2\alpha}(j) \\
& +J_{\perp} \sum_{i} \bm{S}_{x^21}(i) \cdot \bm{S}_{x^22}(i),\\
H_{J,z^2}&=J_{\perp} \sum_{i} \bm{S}_{z^21}(i) \cdot \bm{S}_{z^22}(i).
\label{eq:xz-t-H}
\end{aligned}
\end{equation}
The $3d_{x^2-y^2}$-orbital $H_{J,x^2}$ takes the form of a bilayer $t$-$J_{\parallel}$-$J_{\perp}$ model \cite{matx2022,lu2023bilayertJ}.
Similar single-orbital bilayer $t$-$J_{\parallel}$-$J_{\perp}$ models have also been considered in the context of cuprates ~\cite{ubbens1994,kuboki1995,maly1996,nazarenko1996bilayer3d,medhi2009,eder1995,vojta1999,zhao2005,zegrodnik2017,matx2022} and in cold atom systems~\cite{bohrdt2021exploration,demler2022,Grusdt2023}.
For the $3d_{z^2}$-orbital $H_{z^2}$, only the inter-layer superexchange $J_{\perp}$ is relevant.

We adopt the hopping parameters from the DFT calculation provided by Ref. ~\cite{zhang2023trends} for
the LNO material. 
The relevant parameters are taken as, 
$t^{x}_{\parallel}=0.526$ eV, $t^{z}_{\parallel}=0.113$ eV, $t^{z}_{\perp}=0.676$ eV, $t^{xz}_{\parallel}=0.25$ eV,  $\Delta_{g}=0.528$ eV.
The superexchange coupling parameters $J_\parallel$ and $J_\perp$ should in principle be determined by various interaction parameters in the strong-coupling limit. 
For simplicity, we use the simple relation $J\approx 4t^2/U$ by neglecting other interaction parameters.
Since the accurate value of the Hubbard $U$ is hard to obtain, $J_\parallel$ and $J_\perp$ are set as tuning parameters. 
Since $J_\perp/J_\parallel\approx (t^{z}_{\perp}/t^{x}_{\parallel})^2$, from the tight-binding parameters provided in Ref.~\cite{zhang2023trends} and Ref.~\cite{YaoDX2023}, the ratio between two superexchange strengths is approximated as $J_\perp/J_\parallel\approx 1.7-1.8$. 
Similarly,  $J_\perp/J_\parallel\approx 2$ as given in 
Ref.~\cite{WuWei2023charge}. 
In our calculation, their values are tuned with
their ratio fixed as $J_\perp/J_\parallel=2$ for convenience. 
An estimation in Ref.~\cite{pardo2011dft} provides $U=4.75$ eV, leading to $J_\parallel\approx0.23$ eV.
On the other hand, a smaller value of $J_\parallel=0.1$eV  is used Ref.~\cite{WuWei2023charge}.
Considering the different values of parameters in literature, 
$J_\parallel$ is varied from 0.1 to 0.3eV in our calculation to cover these variations.

\section{The Slave-boson-mean-field theory and band renormalization}
\label{sec:SBMF_Band}

\subsection{The Slave-boson-mean-field (SBMF) Approach}
In the slave boson approach ~\cite{kotliar1988,lee2006htsc}, 
the electron operators of two $E_g$ orbitals are represented as $d_{a\alpha\sigma}^{\dagger}(i)=f_{a\alpha\sigma}^{\dagger} (i) b_{a\alpha}(i)$, where $f_{a\alpha\sigma}^{\dagger} (i)$ is the spinon creation operator and $b_{a\alpha}(i)$ is the holon annihilation operator for the two orbitals ($a=x^2,z^2$) respectively.
The following no-double-occupancy constraint is imposed on each orbital at each site,
\begin{equation}
\begin{aligned}
\sum_{\sigma} f_{a\alpha\sigma}^{\dagger} (i) f_{a\alpha\sigma} (i)
+b_{a\alpha}^{\dagger}(i) b_{a\alpha}(i)=1.
\end{aligned} 
\end{equation}
These constraints are imposed through Lagrange multipliers $\lambda_{a\alpha} (i)$ and in the mean-field (MF) theory, $\lambda_{a\alpha}(i)\equiv \lambda_{a}$ on average irrespective of the layer and site.

The total electron filling $x_{\text{tot}}$ of the $E_g$ orbital is determined from the material, which yeids
$x_{\text{tot}}=0.75$ for LNO.
The NN hybridization causes the density fluctuation between two orbitals while keeps the total holon density $\delta_{\text{tot}}=2-2x_{\text{tot}}$ 
in the $E_g$ orbitals per Ni site fixed,
\begin{align}\label{total_hole}
\frac{1}{N} \sum_{i\alpha} 
\big(b_{x^2\alpha}^{\dagger} (i) b_{x^2\alpha} (i)
+b_{z^2\alpha}^{\dagger} (i) b_{z^2\alpha} (i)
\big) =2\delta_{\text{tot}}.
\end{align}
For the case of undoped LNO, $\delta_{\text{tot}}=0.5$. 
Upon chemical doping, 
$\delta_{\text{tot}}$ varies within $0\sim 1$, and 
is controlled by $\lambda_{\text{tot}}$.
$\lambda_{\text{tot}}$ plays the role of chemical potential.
In addition to $\lambda_{\text{tot}}$, $\lambda_{x^2}$, $\lambda_{z^2}$ are also introduced.
Their combinations together with the onsite 
$\Delta_g$ act as 4 chemical potentials for the four degrees of freedom, i.e., spinons and holons in two 
$E_g$ orbitals, respectively.

After the decomposition, the neareast neighboring AFM superexchange $J_\parallel$ and $J_\perp$ terms give rise to spinon's intra-orbital bondings and singlet pairings.
The $\chi_{a}^{(\alpha)}(i,j)$ ($\alpha=1,2$) and $\chi_{a }^{\perp}(i)$ represent spinon's intra- and inter-layer hopping, respectively. $\Delta_{a}^{(\alpha)}(i,j)$  and $\Delta_{a}^{\perp}(i)$ represent the intra- and inter-layer singlet pairings, respectively.
The electron hopping terms of $t^{a}_{\parallel}$ and $t^{z}_{\perp}$ are decoupled as the intra-orbital spinon hopping $\chi_{a}^{(\alpha)}(i,j)$ and $\chi_{z}^{\perp}(i)$ as well as holon hopping $\kappa_{a}^{(\alpha)}(i,j)$ and $\kappa_{z}^{\perp}(i)$, respectively.
The inter-orbital electron hybridization $t_{\parallel,j-i}^{xz}$  term is decoupled as the inter-orbital spinon hybridization $\chi_{xz}^{(\alpha)}(i,j)$ and holon hybridization $\kappa_{xz}^{(\alpha)}(i,j)$ ($\alpha=1,2$),
respectively.

\begin{table}[t]
\centering
\begin{TAB}(r,0.05cm,0.1cm)[2pt]{|c|c|c|c|c|}{|c|c|c|c|c|c|c|c|c|}
OP &  GS value (eV) & & 
OP &  GS value (eV) \\
$\chi_{\parallel}^{x}$  & $3.1\times 10^{-2}$
& & $\chi_{\parallel}^{z}$ &$-1.6\times 10^{-6}$ \\
$\chi^{x}_{\perp}$  & $4.2\times 10^{-4}$
& & $\chi^{z}_{\perp}$ & $1.8\times 10^{-1}$ \\
$\Delta_{\parallel}^{x}$  & $2.2\times 10^{-5}$ &
& $\chi_{\parallel}^{xz}$ & $4.5\times 10^{-3}$
 \\
$\Delta^{x}_{\perp}$  & $9.7\times 10^{-3}$
& & $\Delta^{z}_{\perp}$ & $4.0\times 10^{-2}$  \\
$\kappa_{\parallel}^{x}$ & $2.5\times 10^{-1}$
& & $\kappa^{z}_{\parallel}$ & $2.2\times 10^{-3}$ \\
$\kappa_{\parallel}^{xz}$ & $2.4\times 10^{-2}$
& & $\kappa^{z}_{\perp}$  &$1.3\times 10^{-2}$ \\
 &  Holon density
& &  & Holon density \\
$\delta_{x^2}$ &  $4.8\times 10^{-1}$
& & $\delta_{z^2}$ & $2.0\times 10^{-2}$ \\
\end{TAB}
\caption{Table of the hopping and pairing order parameters (OPs) as well as the holon densities $\delta_{x^2}$ and $\delta_{z^2}$ calculated by the SBMF theory in the ground state (GS).
The corresponding band and interaction parameter values are  $t^{x}_{\parallel}=0.526$ eV, $t^{z}_{\parallel}=0.113$ eV, $t^{z}_{\perp}=0.676$ eV, $t^{xz}_{\parallel}=0.25$ eV,  $\Delta_{g}=0.528$ eV,
$J_{\parallel}=0.25$eV, and
$J_{\perp}/J_{\parallel}=2$.
}
\label{tab:MForderParameter}
\end{table}

The various parameters defined above are expressed in terms of the expectation values of the corresponding fermion bi-linear operators at thermal equilibrium.
At the mean-field level, order parameters are assumed to be site/layer independent.
The spinon hopping parameters are expressed as
\begin{equation}
\begin{aligned}\label{self_consistent}
\chi_{\parallel}^{a}
&=\frac{3}{8} J_{\parallel} \langle f_{a\alpha\uparrow}^{\dagger}(j) f_{a\alpha\uparrow}(i)
+f_{a\alpha\downarrow}^{\dagger}(j) f_{a\alpha\downarrow}(i)\rangle ,   \\
\chi^{a}_{\perp}
&=\frac{3}{8} J_{\perp} \langle f_{a2\uparrow}^{\dagger}(i) f_{a1\uparrow}(i)
+f_{a2\downarrow}^{\dagger}(i) f_{a1\downarrow}(i)\rangle, \\
\chi_{\parallel}^{xz} 
&=t_{\parallel}^{xz} \langle f_{z^2\alpha\uparrow}^{\dagger}(j) f_{x^2\alpha\uparrow}(i)
+f_{z^2j\alpha\downarrow}^{\dagger}(j) f_{x^2\alpha\downarrow}(i)\rangle,
\end{aligned}
\end{equation}
respectively, where $\langle \cdots \rangle$ means the thermal average.
The spinon pairing order parameters are expressed 
as
\begin{equation}
\begin{aligned}\label{self_consistent2}
\Delta_{\parallel}^{a}
&=\frac{3}{8} J_{\parallel}\langle f_{a\alpha\downarrow}(j) f_{a\alpha\uparrow} (i)
-f_{a\alpha\uparrow}(j) f_{a\alpha\downarrow}(i) \rangle,  \\
\Delta^{a}_{\perp}
&= \frac{3}{8} J_{\perp} \langle f_{a2\downarrow}(i) f_{a1\uparrow}(i) 
-f_{a2\uparrow}(i) f_{a1\downarrow}(i) \rangle,  
\end{aligned}
\end{equation}
respectively.
Similarly for the holon part, their hopping parameters are defined as follows as
\begin{equation}
\begin{aligned}
\kappa_{\parallel}^{a}
=& t_{\parallel}^a\langle b_{a\alpha}^{\dagger}(j) b_{a\alpha}(i) \rangle,  \\
\kappa^{a}_{\perp}=& t_{\perp}^a\langle b_{a2}^{\dagger}(i) b_{a1}(i) \rangle, \\
\kappa_{\parallel}^{xz}= & t_{\parallel}^{xz}\langle b_{x^2\alpha}^{\dagger}(j) b_{z^2\alpha}(i) \rangle. \\
\end{aligned}
\end{equation}

The resultant mean-field Hamiltonians are presented in Appendix ~\ref{appendix:sbmft}.
In the numerical calculations, all the order parameters and the Lagrange multipliers are solved self-consistently. 
In the LNO system, the physical occupations of the $3d_{z^2}$ and $3d_{x^2-y^2}$ orbitals slightly deviate from half- and quarter-filling, respectively, caused by their hybridization.  
The filling fraction $x_a$ is related to the corresponding holon density $\delta_{a}$  via
\bea
x_a=(1-\delta_{a})/2,
\eea
in each $E_g$ orbital represented by $a$.

For a typical value of the intra-layer superexchange $J_{\parallel}=0.25$ eV, the order parameters 
are calculated as listed in Table.~\ref{tab:MForderParameter}. 
The inter-layer $s$-wave pairing is stronger than the 
intra-layer one as expected from the larger inter-layer superexchange.
Note that the obtained $\delta_{z^2}=0.02$ suggests 
that the holon density in the $3d_{z^2}$ orbital is very low, which influences the electronic nature of the system 
as will be seen in the following sections.

\subsection{Correlation induced band flattening}
\begin{figure}[t]
\includegraphics[width=1.0\linewidth]{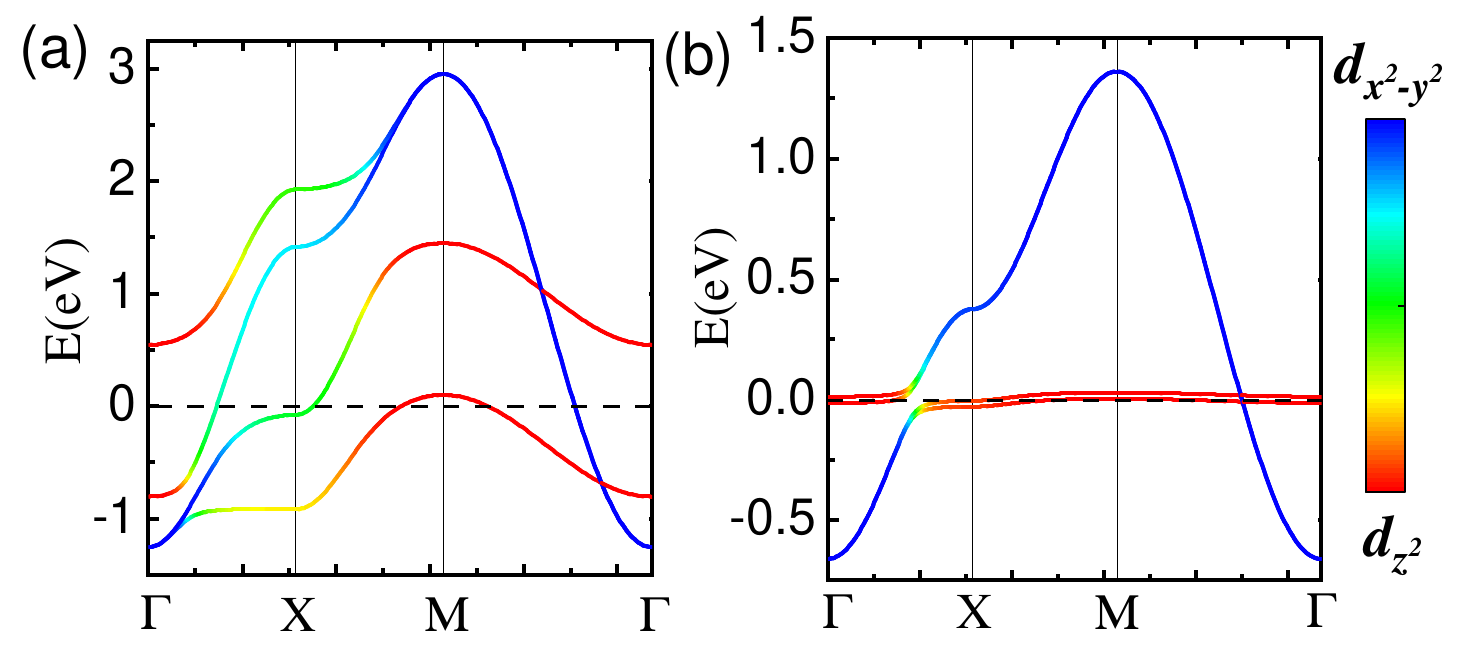}
\caption{($a$) Tight-binding band structure; 
($b$) the spinon band structure. 
The right-side color panel shows the orbital component with the 
red and blue colors representing the $3d_{z^2}$ and $3d_{x^2-y^2}$ orbitals, respectively.
There exist four bands respecting two orbitals and bonding/antibonding hybridization.
In the Mott limit, strong correlation effect flattens 
the the $3d_{z^2}$ orbital dramatically and reduces
the band splitting.
The high symmetry points in the Brillouin zone are
marked as $\Gamma(0,0)$, $X=(\pi,0)$, $M=(\pi, \pi)$.
(Note that the same band structure 
in ($a$) was first published in Ref. \cite{luo2023high}. 
It is presented here for comparison.)
}
\label{fig:BandSpectrum}
\end{figure}

It is important to note that the band structure is significantly renormalized by strong electron correlation here. 
The tight-binding model yields the bare band structure as shown in Fig.~\ref{fig:BandSpectrum} ($a$), and in contrast the spinon band calculated in the SBMF theory is shown Fig.~\ref{fig:BandSpectrum} ($b$). 
The bare band structure shows four bands corresponding to two $E_g$ orbitals and two layers, which are mixed together due to the inter-orbital hybridization and inter-layer hopping.
The lowest band is composed of the inter-layer bonding state of the $3d_{z^2}$ orbital, mixed with  $3d_{x^2-y^2}$ through hybridization.
In contrast, the anti-bonding band of $3d_{z^2}$ is at high energy which can be neglected at low energy.
The other two bands are dominated by the $3d_{x^2-y^2}$-orbital component. 
The above picture results in three Fermi surfaces:
One is hole-like due to the bonding band of the $3d_{z^2}$-orbital, and the other two arise from $3d_{x^2-y^2}$. 

As a comparison, in the spinon band shown in Fig.~\ref{fig:BandSpectrum}($b$), both the $3d_{x^2-y^2}$-orbital bands and the $3d_{z^2}$ bands (the hybridized bands) are significantly flattened due to the constraint of no-double-occupancy. 
The widths of the $3d_{x^2-y^2}$-orbital bands
are reduced by a mild renormalization factor of $\delta_{x^2}\approx 0.5$.
In contrast, the $3d_{z^2}$-band width (the hybridized bands) is much more strongly 
suppressed, renormalized by the factor of the holon densities $\sqrt{\delta_{x^2}\delta_{z^2}}\approx 0.1$
according to the holon densities calculated in Tab. \ref{tab:MForderParameter}.
Here, holon hybridization order parameter $\kappa_{\parallel}^{xz}$ is very small since the $3d_{z^2}$-orbital is nearly half filled with $\delta_{z^2}=0.02$. 
This result is qualitatively consistent with the recent angular-resolved-photo-emission-spectroscopy (ARPES) experiment \cite{yang2023arpes}.

The ARPES detects the spectral function $A(\mathbf{k},\omega)$ of an electron below 
the Fermi level. 
For a free band fermion, $A(\mathbf{k},\omega)=\delta\left(\omega-\varepsilon_\mathbf{k}\right)$, where $\varepsilon_\mathbf{k}$ 
indicates the band energy. 
For the $t$-$J$ model, under the SBMF framework, the $A(\mathbf{k},\omega)$ becomes the convolution of
the spectral functions of a spinon and a holon,
respectively.
Below the holon condensation temperature, $A(\mathbf{k},\omega)$ consists of a low-energy $\delta$-peak of the spinon dispersion
$\varepsilon_\mathbf{k}^f$ and a high-energy incoherent part. 
Above the holon condensation temperature, the $\delta$-peak is broadened, nevertheless, 
the spectral center remains approximately located at $\varepsilon_\mathbf{k}^f$. 
Note that in Fig.~\ref{fig:BandSpectrum}($b$), only the interlayer bonding $3d_{z^2}$ spinon band and the lower half of the $3d_{x^2-y^2}$- spinon bands
are physical, since the decomposition $c_{i\sigma}=f_{i\sigma}b_{i}^\dagger$ only stands when the spinon filling is less than $\frac{1}{2}$.  

When holons are condensed, ARPES measures the spinon spectral function below the Fermi energy. 
Consequently, the ARPES observation (under ambient pressure) yields that the band width of $3d_{z^2-y^2}$ is flattened by a factor of about $\frac{1}{2}$, that of the $3d_{z^2}$ bonding band (the hybridized band) is flattened by a factor of about $\frac{1}{7}\sim \frac{1}{8}$ \cite{yang2023arpes}. 
These observations are well consistent with our result. 
Note that although the ARPES experiment is carried out under ambient pressure, the strong correlation features remain.

\section{Superconductivity}
\label{sec:SC}
There are two aspects regarding to the onset of superconductivity in each orbital, {\it i.e.}, the gap formation and phase coherence. 
In the slave-boson approach, the gap formation is generated by the spinon pairing, and the phase coherence arises from the holon condensation \cite{kotliar1988,lee2006htsc}. 
Consequently, the superconducting order parameter is represented by the product between the spinon and holon parts, 
\begin{equation}
\begin{aligned}\label{superconducting_ODP}
&\Delta^{a}_{\text{SC}}
= \frac{3}{8} J\big\langle c_{a\alpha\downarrow}(j) c_{a\beta\uparrow}(i) 
-c_{a\alpha\uparrow}(j) c_{a\beta\downarrow}(i)  \big\rangle \\
&=\frac{3}{8} J\big\langle f_{a\alpha\downarrow}(j)  f_{a\beta\uparrow}(i)  
-f_{a\alpha\uparrow}(j) f_{a\beta\downarrow}(i)  \big\rangle
\big\langle b_{a\alpha}^{\dagger}(j) b_{a\beta}^{\dagger}(i) \big\rangle.
\end{aligned}
\end{equation}
In the presence of holon condensation, each holon operator is simply replaced by 
a number as
\begin{equation}
b_{a\alpha}^{\dagger}(j) 
= b_{a\alpha}(i) \equiv \sqrt{\delta_{a}}.
\end{equation}
Therefore, the pairing symmetry is determined by the spinon part. 
The numerical results listed in Table.~\ref{tab:MForderParameter} show an extended $s$-wave pairing for both orbitals. 
The inter-layer pairing strength is much stronger than the intra-layer one.  

The driving force of this extended $s$-wave pairing is the strong interlayer superexchange interaction $J_{\perp}$ \cite{matx2022,lu2023bilayertJ}. 
The superconducting $T_c$ is determined by four temperature scales, including two spinon pairing temperatures $T_{\text{pair}}^{x}$ and $T_{\text{pair}}^{z}$, and two holon condensation temperatures $T_{\text{BEC}}^{x}$ and $T_{\text{BEC}}^{z}$. 
In two dimensions, the holon condensation temperature could be estimated from the Kosterlitz-Thouless (KT) transition point \cite{kosterlitz1973kt}. 
In the following, we shall first study the superconductivity in each orbital separately, and the coupling between them. 

\subsection{The $3d_{z^2}$ orbital induced SC}
\begin{figure}[t]
\includegraphics[width=0.8\linewidth]{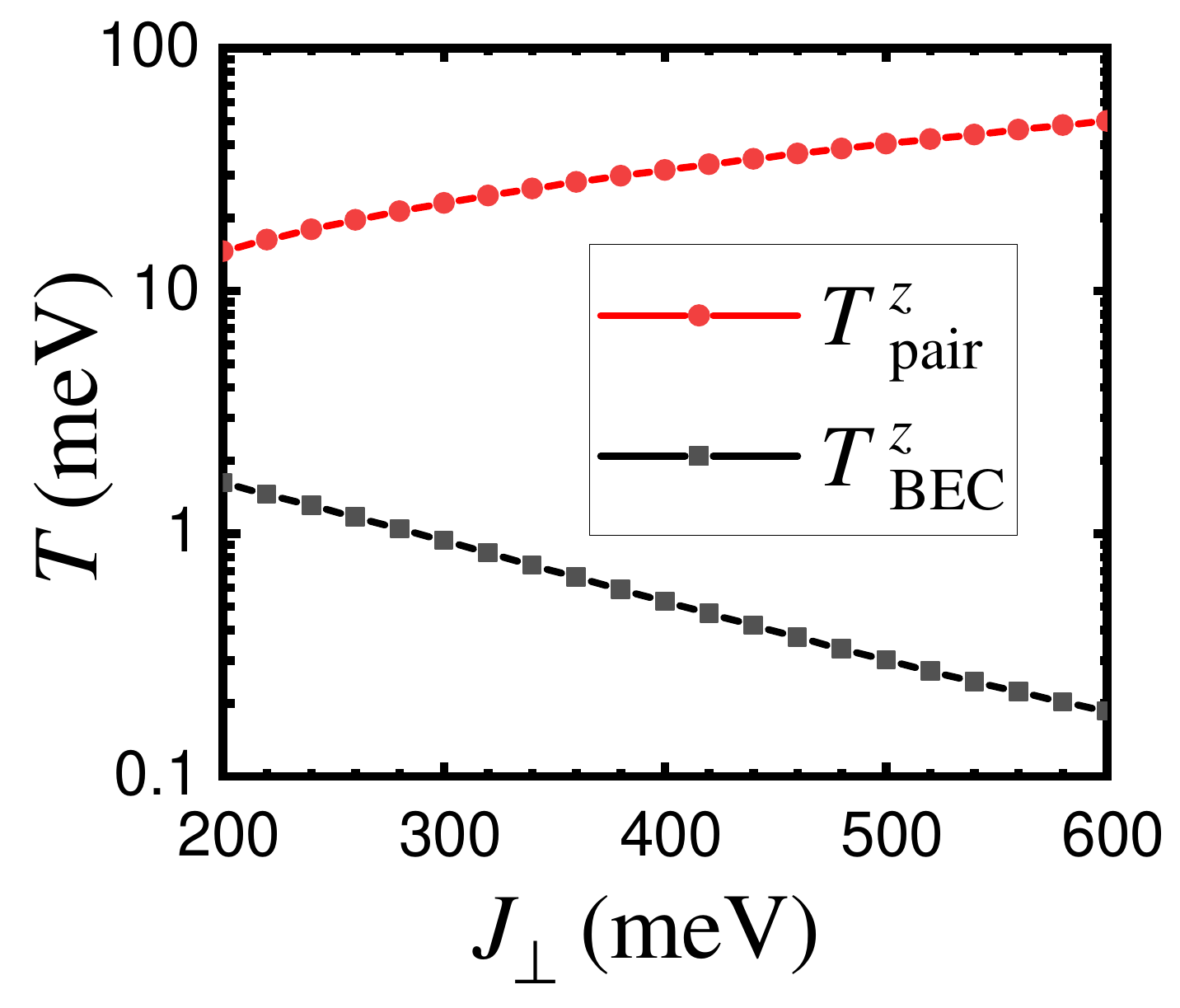}
\caption{Holon condensation temperature $T_{\text{BEC}}^{z}$ and spinon pairing temperature $T_{\text{pair}}^{z}$ versus inter-layer superexchange $J_{\perp}$ for the $3d_{z^2}$ orbital. The on-set temperature of the $3d_{z^2}$ superconductivity is controlled by $T_{\text{BEC}}^{z}$, which is much smaller.
than $T_{\text{pair}}^z$.}
\label{fig:JpTcZ2}
\end{figure}

As shown in Table.~\ref{tab:MForderParameter}, 
the ground-state value of $\Delta^{z}_{\perp}$ is largest among the pairing and hopping order parameters, which is brought by the strong inter-layer superexchange $J_{\perp}$ and the flatness of the $3d_{z^2}$-orbital band (i.e. the hybridized band). 
Consequently, the $3d_{z^2}$ orbital spinons start to pair at a high temperature $T_{\text{pair}}^z$ as shown in Fig.~\ref{fig:JpTcZ2}. 

However, to reach superconductivity in the $3d_{z^2}$ orbital, the $3d_{z^2}$ holons need to condense. 
In 2D, the holon BEC is viewed as the KT transition which requires a finite holon density and the in-plane coherence. 
If without the orbital hybridization, the $3d_{z^2}$ orbital is exactly half-filled and there is no condensation. 
Taking into account hybridization, the $3d_{z^2}$ orbital acquires a finite hole doping transferred from $3d_{x^2-y^2}$. 
Moreover, the original localized $3d_{z^2}$ orbital acquires a narrow band width.
Our numerical results displayed in Table.~\ref{tab:MForderParameter} verify this scenario.

The holon Hamiltonian is presented in Eq.~(\ref{H_holon}) in Appendix \ref{appendix:sbmft}. 
Treating the hybridization term $H_{xz}^{(b)}$ as a perturbation, we investigate how it affects the in-plane hopping of the $3d_{z^2}$- holon. 
This can be done by directly diagonalizing the Hamiltonian Eq.~(\ref{H_holon}), and extracting the band dispersion. 
Consequently, the effective holon Hamiltonian for 
the $3d_{z^2}$ orbital is given by, 
\begin{equation}
\label{effective_holon_H}
\begin{aligned}
\tilde{H}_{z^2}^{(b)}
=&\sum_{\bm{k}\alpha} \tilde{\omega}_{z^2}(\bm{k}) b_{z^2\alpha\bm{k}}^{\dagger} b_{z^2\alpha\bm{k}}     \\
=&\sum_{i,j,\alpha} \tilde{t}_{j-i} b_{z^2\alpha}^{\dagger}(i) b_{z^2\alpha}(j),
\end{aligned}
\end{equation}
where $\tilde{\omega}_{z^2}(\bm{k})$ is the renormalized holon dispersion and
$\tilde{t}_{j-i}$ reflects the effective in-plane hopping integral between the $i$ and $j$ sites. 
For a large distance $l$, the effective in-plane hopping integral decays promptly. 
In the numerical simulation, we only take the four nearest distances, as summarized in Table.~\ref{tab:HybriHopping}. 
The displayed effective in-plane hopping integrals in Table.~\ref{tab:HybriHopping} ($\sim 0.1$meV) are much smaller compared to the typical bare hopping integral of the system ($t_{\parallel}^z,t_{\parallel}^{xz}\sim \mathcal{O}(0.1)$eV). 
The reason for this result lies in the following.

The effective in-plane hopping of the $3d_{z^2}$-orbital holons arises from the direct NN hopping and the indirect process intermediated by the hybridization with the $3d_{x^2-y^2}$ orbital. 
The bare hopping integrals of these two process are not so small: $t_{\parallel}^z=0.113$eV and $t_{\parallel}^{xz}=0.25$eV.
Nonetheless, under the no-double-occupancy constraint the in-plane holon hopping integrals should be renormalized by the spinon order parameters $\chi_{\parallel}^z$ and $\chi_{\parallel}^{xz}$ for the direct hopping and hybridization respectively, as shown in Eq. (\ref{H_holon}).
These renormalization factors are reversely determined by the spinon Hamiltonian Eq. (\ref{H_spinon}) in Appendix \ref{appendix:sbmft}. 
Due to the low holon density of the $3d_{z^2}$-orbital, its effective NN hopping strength $\chi_{\parallel}^z=-1.6\times 10^{-3}$meV is extremely small.

On the contrary, the spinon hybridization order parameter $\chi_{\parallel}^{xz}=4.5$meV is not very small. 
However, the hybridization-intermediated effective hopping for the $3d_{z^2}$ holons is only a second-order perturbative process, which leads to an effective hopping integral roughly estimated as $\tilde{t}\sim (\chi_{\parallel}^{xz})^2/1\text{eV} \sim 0.1$meV. This accounts for the small effective hopping integrals in Table.~\ref{tab:HybriHopping}.

The effective holon model Eq.~(\ref{effective_holon_H}) undergoes KT transition due to phase fluctuations. 
Under the replacement, $b_{z^2\alpha}^{\dagger}(i) =\sqrt{\delta_{z^2}} e^{i\theta_{z^2\alpha}(i)}$, the system reduces to a generalized 2D XY-like model with interactions determined by the effective hopping integrals $\tilde{t}_{j-i}$.
The superfluid stiffness $\rho_{z^2,s}$ could be further estimated from $\tilde{t}_{j-i}$,
\begin{equation}
\rho_{z^2,s}= 2 \delta_{z^2} \sum_{\vec{l}} \tilde{t}_{\vec{l}}\, l^2,
\end{equation}
where the summation takes over the bond directions $\vec{l}$.
The KT transition temperature is proportional to the superfluid stiffness, $T_{\text{BEC}}^{z}=\frac{\pi}{2}\rho_{z^2,s}$ \cite{kosterlitz1973kt}.

\begin{table}[t]
\centering
\begin{tabular}{|c|c|c|}
\hline
\makecell{direction \\ vector $\vec{l}$ } & distance $l$ & \makecell{hopping integral \\  $\tilde{t}_{\vec{l}}$ (meV)}  \\
\hline
$(1,0)$ & $1$ & $0.167$  \\
\hline
$(1,1)$ & $\sqrt{2}$ & $0.726$  \\
\hline
$(2,0)$ & $2$ & $0.580$  \\
\hline
$(2,1)$ & $\sqrt{5}$ & $0.212$  \\
\hline
\end{tabular}
\caption{Effective hopping integrals $\tilde{t}_l$ for the $3d_{z^2}$-orbital holon at $J_{\perp}/J_{\parallel}=2$, $J_{\perp}=0.5$eV without doping.}
\label{tab:HybriHopping}
\end{table}

The simulated pairing temperature $T_{\text{pair}}^{z}$ and phase coherence temperature $T_{\text{BEC}}^{z}$ as function of $J_{\perp}$ are displayed in Fig.~\ref{fig:JpTcZ2}. 
For the $3d_{z^2}$ orbital, $T_{\text{pair}}^{z}$ is larger than $T_{\text{BEC}}^{z}$ in several orders of magnitude.
The SC of $3d_{z^2}$ orbital is determined by the coherence temperature $T_{\text{BEC}}^{z}$. 
As $J_{\perp}=2J_{\parallel}$ increases, the inter-layer pairing of the $3d_{z^2}$ orbital is enhanced simultaneously, leading to a stronger $T_{\text{pair}}^{z}$.
On the contrary, the small $T_{\text{BEC}}^{z}$ originates from the lack of holon density $\delta_{z^2}$ near half filling as well as the perturbative generation of the hopping integrals $\tilde{t}$, both of which are very small. 
As a result, the $T_{\text{BEC}}^{z}$ for realistic $J_{\perp}$ is much lower than the experimental $T_c\approx 80$ K in LNO \cite{Wang2023LNO}. 
The experimental observed high superconducting $T_c$ may not be caused by the $3d_{z^2}$-orbital due to the low  $T_{\text{BEC}}^{z}$.

\subsection{The $3d_{x^2-y^2}$ orbital induced SC}
The SC in the $3d_{x^2-y^2}$-orbital is analyzed by the same approach. 
Previously, we have derived an effective single $3d_{x^2-y^2}$-orbital bilayer $t$-$J$ model by integrating out the $3d_{z^2}$-orbital spin degree of freedom~\cite{lu2023bilayertJ}, 
in which the $J$-term is just represented by the $H_{x^2}$ part of Hamiltonian in Eq. (\ref{eq:xz-t-H}). 
For the $3d_{x^2-y^2}$-orbital, the hybridization term 
in Eq. \ref{eq:ham0}
is only a perturbation, which does not obviously affect $T_c$. 
Therefore, this term  is first neglected in this subsection, and then will be investigated
in detail in next subsection.

\begin{figure}[t]
\includegraphics[width=0.8\linewidth]{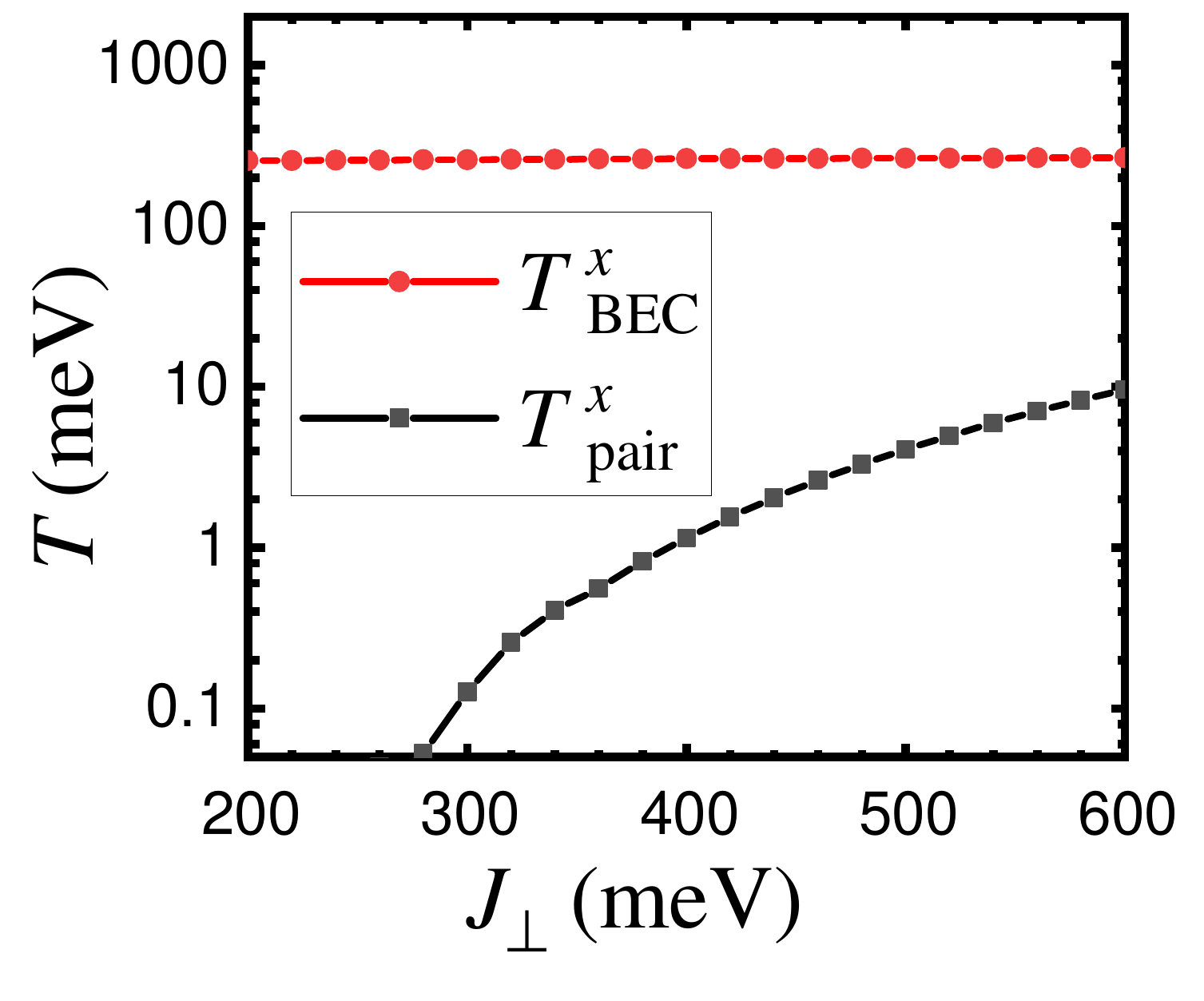}
\caption{Holon condensation temperature $T_{\text{BEC}}^{x}$ and spinon pairing temperature $T_{\text{pair}}^{x}$ versus inter-layer superexchange $J_{\perp}$ for the $3d_{x^2}$ orbital. 
}
\label{fig:JpTcX2}
\end{figure}

The condensation temperature of the $3d_{x^2-y^2}$-orbital holons $T_{\text{BEC}}^x$ shown in Fig.~\ref{fig:JpTcX2} is very high ($\approx 200$ meV $\approx 2000$ K), which serves as the largest temperature scale among the four temperatures relevant to the SC. 
Why $T_{\text{BEC}}^x$ is so high is due to the following two aspects: the high holon density
{\it i.e.}, 0.5, and the large in-plane hopping integral of the $3d_{x^2-y^2}$-orbital.
This sets up the coherence condition for the $3d_{x^2-y^2}$-orbital to exhibit high-$T_c$ 
superconductivity.

The pairing temperature of the $3d_{x^2-y^2}$-orbital spinons $T_{\text{pair}}^x$ shown in Fig.~\ref{fig:JpTcX2} arises promptly with the enhancement of $J_{\perp}$ for fixed $J_{\perp}/J_{\parallel}=2$. 
As clarified in our previous work reported in Ref.~\cite{lu2023bilayertJ}, the high $T_{\text{pair}}^x$ originates from the strong interlayer superexchange interaction $J_{\perp}$, which drives the interlayer $s$-wave pairing symmetry in the system. 
As Fig.~\ref{fig:JpTcX2} shows $T_{\text{BEC}}^x\gg T_{\text{pair}}^x$, superconductivity in the $3d_{x^2-y^2}$ orbital degree of freedom is thus determined by its pairing temperature $T_{\text{pair}}^x$. 

\subsection{SC in the two-orbital system}
\begin{figure}[t]
\includegraphics[width=1.0\linewidth]{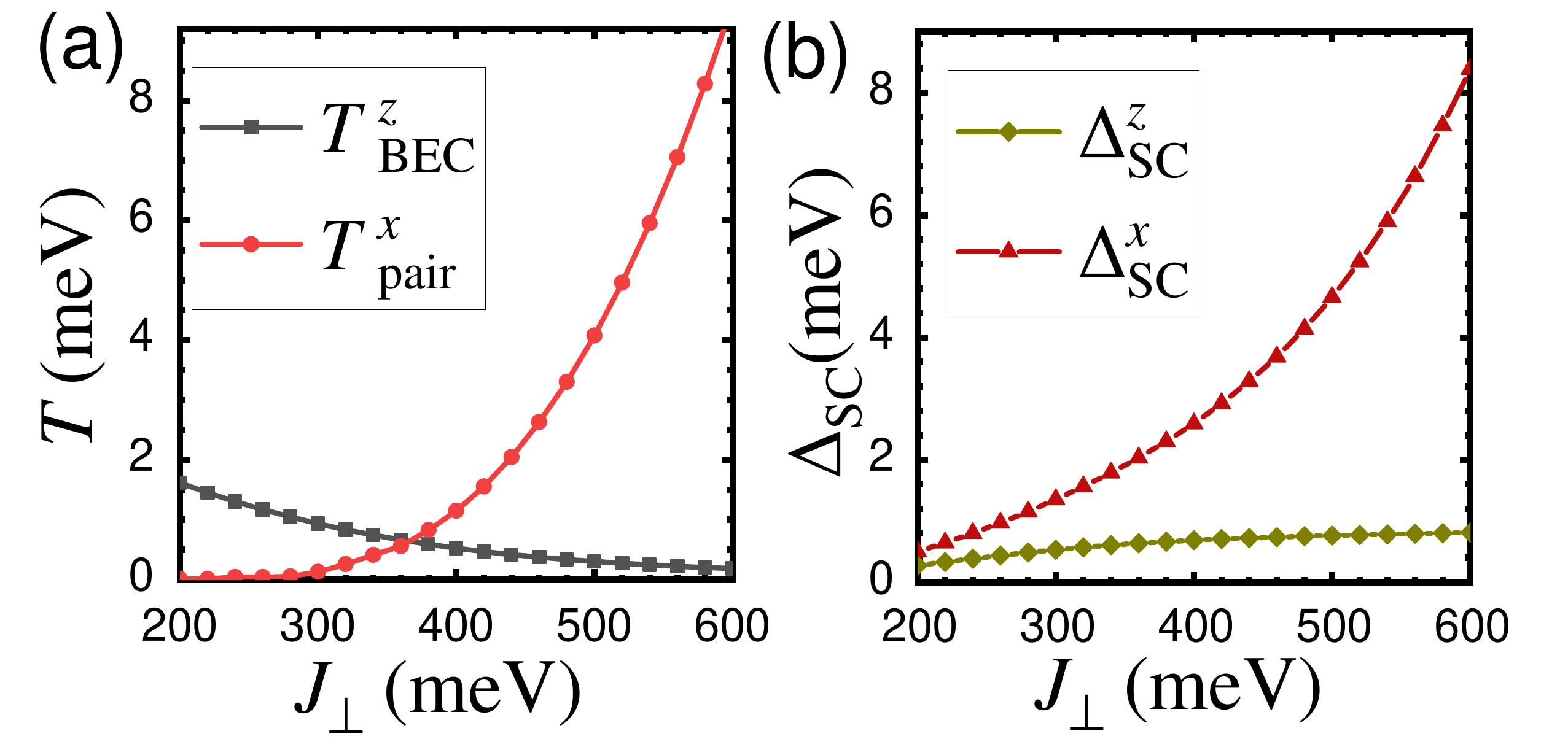}
\caption{(a). Superconducting temperature $T_{\text{BEC}}^{z}$ and $T_{\text{pair}}^{x}$ for the two orbitals. As $J_{\perp}$ becomes stronger, $T_{\text{BEC}}^{z}$ decreases and $T_{\text{pair}}^{x}$ increases.
The superconducting nature will reverse at a critical strength.
(b). Inter-layer superconducting order parameters $\Delta_{\text{SC}}^{z}$ and $\Delta_{\text{SC}}^{x}$ as $J_{\perp}$ varies.
}
\label{fig:TcSCJ}
\end{figure}

As clarified above, while the superconductivity in the $3d_{z^2}$-orbital is determined by its holon BEC temperature $T_{\text{BEC}}^z$, that in the $3d_{x^2-y^2}$-orbital is determined by its spinon pairing temperature $T_{\text{pair}}^x$. 
The dependence of these two temperature scales
on $J_{\perp}$ is shown in Fig.~\ref{fig:TcSCJ}(a) with the fixed
value of $J_{\perp}/J_{\parallel}=2$.
It suggests that $T_{\text{BEC}}^z>T_{\text{pair}}^x$,
or, $T_{\text{BEC}}^z<T_{\text{pair}}^x$, 
at weak or strong superexchange interaction,
respectively.
Since two orbitals are hybridized, when one orbital is superconducting, it induced superconductivity
in the other via the proximity effect.
Therefore, the superconducting $T_c$ in LNO is determined by the higher one between $T_{\text{BEC}}^z$ and $T_{\text{pair}}^x$. 

In the realistic range of $J_{\perp}\in (200, 600)$ meV, Fig.~\ref{fig:TcSCJ}($a$) suggests that $T_{\text{BEC}}^z$ is much lower than the experimental $T_c\approx 80$ K. 
The low $T_{\text{BEC}}^z$ originates from the no-double-occupancy constraint since the hybridization between two $E_g$ orbitals is suppressed by strong correlations.
Consequently, not only the holon density of the $3d_{z^2}$-orbital is very low, but also the effective in-plane holon hopping integrals are weak. 
The combination of these two aspects strongly suppresses the phase coherent temperature of the $3d_{z^2}$ orbitals and hence its superconducting $T_c$. 
What is more, as reported in Ref.\cite{cao2023flat}, the Hund's rule coupling also strongly suppress the hybridization between the two $E_g$ orbitals, which has not been considered here. 
Such an effect will further suppress the $T_c$ of the $3d_{z^2}$ orbital. 
Therefore, we conclude here that the $3d_{z^2}$ orbital 
is unlikely to play a dominant role in providing the superconducting mechanism in the LNO system. 

The $3d_{x^2-y^2}$ orbital instead provides a possible mechanism to
the high $T_c$ in LNO comparable with experiment. 
Fig.~\ref{fig:TcSCJ}(a) suggests that the $T_{\text{pair}}^x$ arises promptly with the enhancement of $J_{\perp}$, which can touch or even surpass the experimentally reported $T_c\approx 80$ K when 
$J_{\parallel}\in \left(0.25, 0.3\right)$ eV. 
Under the relation $J_{\parallel}=4(t_{\parallel}^x)^2/U$, the estimated Hubbard $U=4.75$ eV \cite{pardo2011dft} in combination with the provided $t_{\parallel}^x=0.526$ eV yields $J_{\parallel}=0.23$ eV, which is already very 
close to this regime.
This result suggests that the $3d_{x^2-y^2}$ orbital plays a dominant role in the superconducting mechanism in the LNO system.

It is interesting to investigate the proximity effect between pairing gaps of two orbitals, which is induced by the $H_{xz}$ term in Eq. (\ref{eq:xz-t-H}) neglected in the last subsection. 
Tab.~\ref{tab:MForderParameter} suggests that the $3d_{z^2}$ orbital exhibits a large spinon pairing gap in the ground state, as verified in its large spinon pairing temperature $T_{\text{pair}}^z$ shown in Fig.~\ref{fig:JpTcZ2}. 
This large $3d_{z^2}$-orbital spinon pairing gap considerably enhances the $3d_{x^2-y^2}$-orbital spinon pairing gap through the proximity effect induced by the hybridization between two orbitals since $\kappa^{xz}_{\parallel}$ is nonzero. 

However, this can only happen below the condensation temperature $T_{\text{BEC}}^z$ of the $3d_{z^2}$ orbital, under which SC is achieved in this orbital. 
This is because that, above $T_{\text{BEC}}^z$, the phase fluctuation dictates $\left\langle b_{z^2i\alpha}\right\rangle=0$ such that $\kappa^{xz}_{\parallel}= t^{xz}_{\parallel} \left\langle b_{x^2i\alpha}^\dagger b_{z^2i\alpha}\right\rangle=0$. 
This analysis suggests that in Fig.~\ref{fig:TcSCJ} ($a$),
in the regime with $T_{\text{BEC}}^z<T_{\text{pair}}^x$, the proximity cannot happen when the temperature is near $T_{\text{pair}}^x$, and thus $T_c$ is set as $T_c=T_{\text{pair}}^x$ which is blind to the $H_{xz}$ term in Eq. (\ref{eq:xz-t-H}).
In contrast, in the regime with $T_{\text{BEC}}^z>T_{\text{pair}}^x$, the proximity effect directly drives superconductivity in the $3d_{x^2-y^2}$ orbital when $T<T_{\text{BEC}}^z$. 
This analysis suggests that the physical $T_c$ is determined by the higher one of $T_{\text{BEC}}^z$ and $T_{\text{pair}}^x$ shown in Fig.~\ref{fig:TcSCJ}(a), 
which is not affected by the $H_{xz}$ term 
in Eq. (\ref{eq:xz-t-H}) in the SBMF framework.

Combining the spinon pairing and holon condensation together, we calculate the physical superconducting order parameter expressed in the form of the $c$-operator via Eq. (\ref{superconducting_ODP}). 
The ground-state interlayer-pairing superconducting order parameter of the two $E_g$ orbitals as function of $J_{\perp}$ is shown in Fig.~\ref{fig:TcSCJ}($b$). 
Clearly, the ground-state superconducting order of 
the $3d_{x^2-y^2}$ orbital is stronger than that of the $3d_{z^2}$ orbital for all values of $J_\perp$ studied although the spinon gap of the former is much weaker than that of the latter, because the holon density in the $3d_{z^2}$ orbital is too low. 

The above result is consistent with that obtained via the dynamic mean-field theory shown in Ref. \cite{tian2023correlation}. 
It is interesting to note that in the regime wherein $T_{\text{BEC}}^z>T_{\text{pair}}^x$ so that the SC is induced by the $3d_{z^2}$ orbital, the ground-state superconducting order of the $3d_{x^2-y^2}$ orbital is still stronger than that of the $3d_{z^2}$ orbital, as the $3d_{x^2-y^2}$ spinon gap is considerably enhanced by the proximity effect in the ground state. 

The viewpoint proposed here on why the high-$T_c$ superconductivity in the LNO only emerges under pressure is different from that proposed in Ref.\cite{Wang2023LNO}. 
In the latter senario, the role of pressure is to lift upward the $\sigma$-bonding band consisting of the $3d_{z^2}$ orbital to cross the Fermi level and metallize it.
Along this line, Refs.\cite{Yi_Feng2023,qin2023high} 
further develop theories viewing $3d_{z^2}$ orbital as
the driving force of superconductivity.
In contrast, here we propose that the main role of pressure lies in that it enhances the interlayer coupling between the $3d_{z^2}$ orbitals to generate a strong interlayer superexchange interaction $J_{\perp}$,
which is further transmitted to the $3d_{x^2-y^2}$ orbitals via the Hund's-rule coupling.
After the $3d_{x^2-y^2}$ orbitals acquire the strong $J_{\perp}$, $T_c$ is strongly enhanced in the entire system. 

Of course, the metallization of the $3d_{z^2}$ band indirectly enhances the filling fraction of the $3d_{x^2-y^2}$ band through Eq. (\ref{total_hole}) and hence benefits its high $T_c$. 
However, the filling fraction of the $3d_{x^2-y^2}$ band can be enhanced by other approaches such as the chemical doping. 
As will be shown in the next section, the electron doping into LNO renders that the $\sigma$-bonding band is completely buried below the Fermi level, but the superconducting $T_c$ enhances promptly. 
Therefore, the metallization of the $\sigma$-bonding band is not necessary for the high-$T_c$ SC in LNO.

\section{Doping Effect to the Phase diagram}
\label{sec:Doping_Effect}

\subsection{The chemical doping}
Chemical doping serves as an important technique which may induce new physical phenomena. 
Here, we study the chemical doping into LNO to explore how superconductivity and pairing nature vary. 
At least, a possible hole-doping candidate is La$_{3-x}$Sr$_{x}$Ni$_2$O$_7$ ($x>0$), whose electron 
number in the $E_g$ orbitals per Ni atom  becomes $1.5-\frac{x}{2}$.

\begin{figure}[t]
\includegraphics[width=1.0\linewidth]{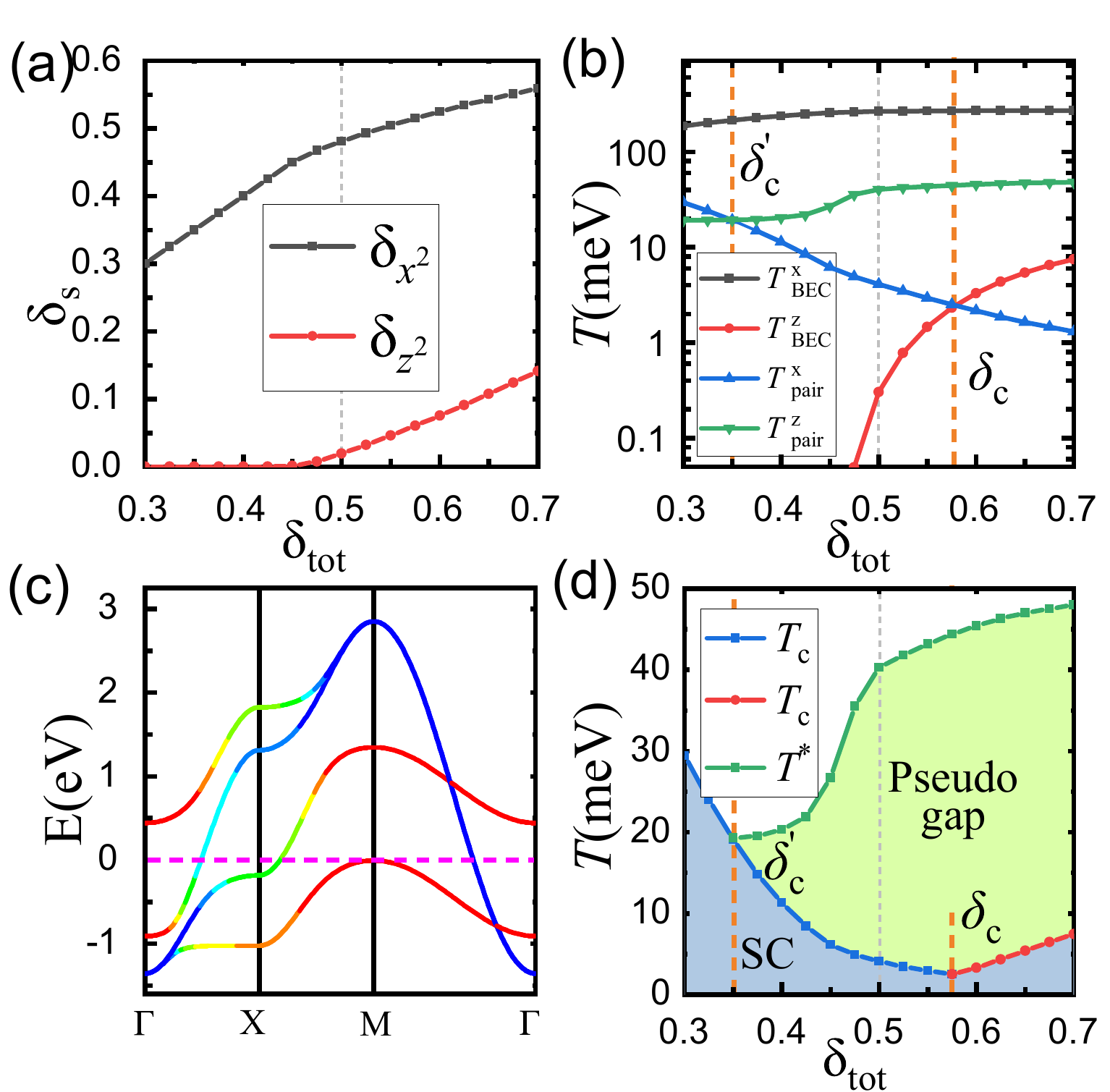}
\caption{(a). Realistic holon densities $\delta_{x^2}$ and $\delta_{z^2}$ versus total holon density $\delta_{\text{tot}}$.
(b). Doping dependence of the relevant four temperature scales for $J_{\parallel}=0.25$eV.
Here, $T_{\text{BEC}}^z$ and $T_{\text{pair}}^x$ cross at a critical doping $\delta_c\approx 0.59$.
$T_{\text{pair}}^z$ and $T_{\text{pair}}^x$ cross at another critical doping $\delta_c^{\prime}\approx 0.35$.
(c). Tight-binding band structure for the holon density $\delta_{\text{tot}}=0.35$.
(d). Superconducting $T_c$ and pseudo gap temperature $T^*$ versus $\delta_{\text{tot}}$.
The upper curve (green line) determines the pseudo gap phase from $3d_{z^2}$ orbital.
The pseudo gap phase merges into the superconducting phase at $\delta_{c}^{\prime}$.
The lower curve determines the onset of superconductivity (SC), which has a local minima at $\delta_c$.
}
\label{fig:dopingQuantity}
\end{figure}

Let us consider by doping the LNO material, such that the total hole number per Ni atom in the $E_g$ orbital is $\delta_{\text{tot}}$.
The derivation of its value from $0.5$ means doping:
$\delta_{\text{tot}}<0.5$, or, $\delta_{\text{tot}}>0.5$ refers to electron-doping or hole-doping the LNO, respectively.
Consequently, both the hole densities $\delta_{s}$ in the two orbitals $(s=x^2, z^2)$ increases as increasing the total hole density $\delta_{\text{tot}}$ as shown in Fig.~\ref{fig:dopingQuantity}($a$).  
The holon condensation temperatures $T_{\text{BEC}}^s$ and spinon pairing temperatures $T_{\text{pair}}^s$ in the two orbitals versus $\delta_{\text{tot}}$ are summarized in Fig.~\ref{fig:dopingQuantity}($b$). 
The relations of the phase coherence and pairing temperatures in each orbital are similar to the undoped case, 
i.e. $\delta_{\text{tot}}=0.5$:
$T_{\text{pair}}^z\gg T_{\text{BEC}}^z$ for the $3d_{z^2}$ orbital, and $T_{\text{BEC}}^x\gg T_{\text{pair}}^x$
for the $3d_{x^2-y^2}$ orbital.
The physical superconducting $T_c$ for the doped LNO is 
the higher of $T_{\text{BEC}}^z$ and $T_{\text{pair}}^x$.

We first consider the case of electron-doping in the regime $\delta_{\text{tot}}<0.5$. 
As shown in Fig.~\ref{fig:dopingQuantity}($b$), $T_{\text{BEC}}^z$ drops to zero promptly
upon electron doping.
The reason lies in that the hole density $\delta_{z^2}$ is suppressed to zero, and then phase coherence in the $3d_{z^2}$-orbital is lost. 
The situation for the $3d_{x^2-y^2}$ orbital is just on the contrary. 
As shown in Fig.~\ref{fig:dopingQuantity}($b$), $T_{\text{pair}}^x$ is quickly enhanced with electron doping, which reduces $\delta_{x^2}$ and enhances the density of state (DOS) of the $3d_{x^2-y^2}$-spinons. 
The similar situation occurs in cuprates: in the heavily over-doped regime, the reduction of doping significantly enhances $T_c$. 
For the physical $T_c$ in LNO under electron-doping, i.e.,$T_c=T_{\text{pair}}^x$, is enhanced with electron doping.

The difference between our viewpoint and that held in Ref.\cite{Wang2023LNO,ZhangGM2023DMRG,Yi_Feng2023,qin2023high} is more obvious in the electron-doped LNO. 
Consider the situation at $\delta_{\text{tot}}=0.35$.
As shown in Fig.~\ref{fig:dopingQuantity}(c), tight-binding calculation yields that the $\sigma$-bonding 
band is completely buried below the Fermi surface.
Then from the latter viewpoint, superconductivity is 
unfavorable. 
However, our result displayed in Fig.~\ref{fig:dopingQuantity} ($b$) suggests that the SC is not only maintained but also enhanced. 

Now we come to the hole doping in the regime $\delta_{\text{tot}}>0.5$. Fig.~\ref{fig:dopingQuantity}($b$) suggests that upon the enhancement of hole doping, $T_{\text{BEC}}^z$ grows promptly, while $T_{\text{pair}}^x$ drops. 
The physical $T_c$ is represents by the higher one between $T_{\text{BEC}}^z$ and $T_{\text{pair}}^x$, as depicted in Fig.~\ref{fig:dopingQuantity}($d$).
At the hole doping level attains a critical one $\delta_c=0.59$, the curves of $T_{\text{BEC}}^z$ and $T_{\text{pair}}^x$ cross.
Thus, in  the higher doping regime $\delta_{\text{tot}}>\delta_c$, $T_c$ is represented by $3d_{z^2}$-holon condensation temperature, while in the lower doping regime $\delta_{\text{tot}}<\delta_c$, $T_c$ is represented by $3d_{x^2-y^2}$-spinon pairing temperature.
$T_c$ shows non-monotonic behavior and has a minimum at $\delta_{\text{tot}}=\delta_c$.

It is interesting that on the two sides of the doping level $\delta_c$, the nature of the superconducting transition is fundamentally different.
Above $\delta_c$, superconducting transition at $T_c=T_{\text{BEC}}^z$ is driven by the $3d_{z^2}$-orbital and belongs to a BEC-type transition.
On the contrary, below $\delta_c$, the transition at $T_c=T_{\text{pair}}^x$ is driven by the $3d_{x^2}$-orbital BCS-type transition.
Such a change can be verified in such experiment as the optical conductivity: for the BEC transition driven by the kinetic energy, the onset of superconductivity is accompanied by a shift of the spectral weight from high to low frequencies; while for the BCS one, such phenomenon does not occur.

\subsection{The Pseudo Gap}
The above two-orbital $t$-$J$ model senario could also yield the pseudogap state in the high T$_c$ LNO.
The pseudogap takes place when spinons are paired while holons are uncondensed in the SBMF picture.
Let us examine two relevant orbitals respectively. 
For the $3d_{x^2-y^2}$ orbital, since 
$T_{\text{BEC}}^x \gg T_{\text{pair}}^x$,
{\it i.e.}, the holon condensation temperature is much higher than the spinon pairing temperature, the pseudogap state should be absent in this 
channel. 
However, the situation is reversed in the $3d_{z^2}$ orbital since 
$T_{\text{BEC}}^z \ll T_{\text{pair}}^z$,
and then $T_{\text{pair}}^z$ behaves as the pseudogap temperature $T^*$. 
Of course, this requires $T_{\text{pair}}^z >T_c$
which is satisfied unless in the extreme 
electron-doping regime. 
In particular, the pseduogap phase should appear
in the undoped LNO, since $T_c=T_{\text{pair}}^x$ and $T^*=T_{\text{pair}}^z$ with $T_c\ll T^*$.
The phase diagram based on the above analysis 
is shown in Fig.~\ref{fig:dopingQuantity}(d), including both the superconducting and pseudo-gap phases.

With the enhancement of hole doping, $T^*=T_{\text{pair}}^z$ increases. 
Consequently, $T^*$ is always higher than $T_c$, even though $T_c$ can also rise when 
$\delta_{\text{tot}}$ is larger than $\delta_c$.
Therefore, the hole-doping regime can host the pseudo gap phase. 
With the enhancement of electron doping, $T^*$ instead decreases due to the reduction of the DOS of the $3d_{z^2}$ spinons, as the Fermi level is lifting up toward the band top. 
Then $T_c=T_{\text{pair}}^x$ on the contrary increases with electron doping due to the enhancement of the DOS of the $3d_{x^2-y^2}$ spinons. 
Consequently, the curves of $T_c$ versus $\delta $
and $T^*$ versus $\delta$ cross at another critical doping $\delta_c^\prime\approx 0.35$. 

Note that the pseudo gap phenomenon here in LNO is significantly different from that in cuprates. 
In cuprates, the ``optimal doping''  divides the phase diagram into the underdoped and overdoped regimes, respectively.
The optimal doping carries a two-fold role: 
It is where $T_c=T^*$, i.e., 
the pseudo-gap phase merges into the superconducting phase; and also where the superconductivity cross over from the BEC type to the BCS one.
However, in LNO, the above two phenomena 
take place at different doping levels:
$\delta_c$ marks the temperature the BEC and the BCS transition curves cross, and 
$\delta_c^\prime$ marks where $T_c=T^*$.
Furthermore, in LNO neither $\delta_c$ nor $\delta_c^\prime$ marks the ``optimal $T_c$'',
and $\delta_c$ actually marks a local minimum of $T_c$. 

\section{Conclusion}
\label{sec:Conclusion}

In this article, we perform SBMF analysis to explore the interplay between $3d_{z^2}$ and $d_{x^2-y^2}$ orbitals based on the bilayer two-orbital $t$-$J$ model. 
Our results reveal that due to the no-double-occupancy constraint, 
the $3d_{x^2-y^2}$ band and the $\sigma$-bonding $3d_{z^2}$ band are flattened by the factor $2$ and $10$ respectively, which is consistent with the ARPES observation \cite{zhou2023evidence}. 
Inter-layer $s$-wave superconducting pairings within the two $E_g$-orbitals dominate in the LNO material. 
For $3d_{z^2}$ orbital, its spinon pairing temperature $T_{\text{pair}}^{z}$ is high and its holon condensation temperature $T_{\text{BEC}}^{z}$ is very low since the no-double-occupancy constraint strongly suppresses the hybridization between the two $E_g$-orbitals.
The superconductivity induced by $3d_{z^2}$ orbital can hardly attain the experimental high $T_c$. 
Instead, the high-$T_c$ superconductivity in LNO is induced by the $3d_{x^2-y^2}$ orbital, whose spinon pairing temperature $T_{\text{pair}}^{x}$ is driven by the strong inter-layer superexchange.

We have also studied the chemical doping effect to the LNO and obtain the phase diagram. 
Under electron doping, $T_c=T_{\text{pair}}^{x}$ continues to increase due to enhancement of the $3d_{x^2-y^2}$-spinons' DOS, although the $\sigma$-bonding $3d_{z^2}$ band is diving below the FS, suggesting that the metallization of this band is not necessary for the high-$T_c$ SC in LNO. 
Under hole doping, $T_c=T_{\text{pair}}^{x}$ initially drops, and above a critical holon density $\delta_c=0.59$ we have $T_c=T_{\text{BEC}}^{z}$,  which begins to rise. 
Across the doping $\delta_c$, the nature of the superconducting transition changes from the BCS type to the BEC type. 
Above the superconducting $T_c$, there exists an extra temperature scale $T^*$, i.e., the pseudo gap temperature, which is given by the spinon pairing temperature of $3d_{z^2}$ orbital. 
The $T^*$ enhances under hole doping and reduces under electron doping. 
The pseudo gap phase merges into the superconducting phase below the holon density $\delta_c^\prime=0.35$. 
This interesting $(T,\delta)$ phase diagram calls for further experimental verification.

\begin{acknowledgements}
We are grateful to the stimulating discussions with Wei Li, Yi-Zhuang You, Wei-Qiang Chen and Meng Wang. F.Y. is supported by the National Natural Science Foundation of China under the Grants No. 12074031, and No. 11674025. C.W. is supported by the National Natural Science Foundation of China under the Grants No. 12234016 and No. 12174317. 
This work has been supported by the New Cornerstone Science Foundation.
\end{acknowledgements}

\appendix
\section{Slave boson mean field theory}
\label{appendix:sbmft}
In this appendix, we perform the details of the slave boson mean field theory in the maintext.
For the inter-layer spin superexchange, the decomposition into hopping and singlet pairing channels is given by,
\begin{equation*}
\begin{aligned}
&J_{\perp} \bm{S}_{ai1} \cdot \bm{S}_{ai2}
=\frac{8}{3J_{\perp}}  \Big[|\chi_{a}^{\perp}(i)|^2 +|\Delta_{a}^{\perp}(i)|^2  \Big] \\
&-\Big[\chi_{a}^{\perp}(i)
\big(f_{a1\uparrow}^{\dagger}(i) f_{a2\uparrow}(i)
+f_{a1\downarrow}^{\dagger}(i) f_{a2\downarrow}(i) \big)
+\text{h.c.} \Big] \\
&-\Big[\Delta_{a}^{(\perp)}(i)
\big(f_{a1\uparrow}^{\dagger}(i) f_{a2\downarrow}^{\dagger}(i) 
-f_{a1\downarrow}^{\dagger}(i) f_{a2\uparrow}^{\dagger}(i) \big) 
+\text{h.c.} \Big], \\
\end{aligned}
\end{equation*}
with $a=x^2,z^2$ for the two orbitals respectively.
Similar decomposition holds for the intra-layer superexchange,
\begin{equation*}
\begin{aligned}
&J_{\parallel} \bm{S}_{a\alpha}(i) \cdot \bm{S}_{a\alpha}(j) 
=\frac{8}{3J_{\parallel}} \Big[|\chi_{a}^{(\alpha)}(i,j)|^2 +|\Delta_{a}^{(\alpha)}(i,j)|^2 \Big] \\
&-\Big[\chi_{a}^{(\alpha)}(i,j)
\big(f_{a\alpha\uparrow}^{\dagger}(i) f_{a\alpha\uparrow}(j)
+f_{a\alpha\downarrow}^{\dagger}(i) f_{a\alpha\downarrow}(j)\big)
+\text{h.c.} \Big] \\
&-\Big[\Delta_a^{(\alpha)}(i,j)
\big(f_{a\alpha\uparrow}^{\dagger}(i) f_{a\alpha\downarrow}^{\dagger}(j) 
-f_{a\alpha\downarrow}^{\dagger}(i) f_{a\alpha\uparrow}^{\dagger}(j) \big) 
+\text{h.c.}  \Big].
\end{aligned}
\end{equation*}
The hopping term can be directly decoupled, e.g.,
\begin{equation*}
\begin{aligned}
&t_{\parallel}^{x} \sum_{\sigma}\big( d_{x^2\alpha\sigma}^{\dagger}(i) d_{x^2\alpha\sigma}(j) +\text{h.c.}\big)   \\
=& \sum_{\sigma}\big( \kappa_{x^2}^{(\alpha)}(i,j) f_{x^2\alpha\sigma}^{\dagger}(i) f_{x^2j\alpha\sigma}(j)+\text{h.c.} \big) \\
& +\big( \frac{t_{\parallel}^{x}}{3J_{\parallel}/8} \chi_{x^2}^{(\alpha)}(i,j) 
b_{x^2\alpha}^{\dagger}(i) b_{x^2\alpha}(j) +\text{h.c.} \big)  
\end{aligned}
\end{equation*}
and similar for the other hopping terms.
Based on these decoupling formulation, we could obtain the mean field Hamiltonian.

After the mean-field decomposition, we arrive at he resultant MF Hamiltonian.
The holon part reads
\begin{widetext}
\begin{equation}
\begin{aligned}\label{H_holon}
&H_{\text{MF}}^{(b)}
=-(\delta\mu+\mu_b) \sum_{i\alpha} 
b_{x^2\alpha}^{\dagger}(i) b_{x^2\alpha}(i) 
-\mu_b \sum_{i\alpha} 
b_{z^2\alpha}^{\dagger}(i) b_{z^2\alpha}(i) \\
&- \frac{t_{\parallel}^z}{3J_{\parallel}/8} \chi_{\parallel}^x \sum_{\langle i,j\rangle \alpha} \big( b_{x^2\alpha}^{\dagger}(i) b_{x^2\alpha}(j)  
+\text{h.c.}\big)  
- \frac{t_{\parallel}^z}{3J_{\parallel}/8} \chi_{\parallel}^z \sum_{\langle i,j\rangle \alpha} \big( b_{z^2\alpha}^{\dagger}(i) b_{z^2\alpha}(j)  
+\text{h.c.}\big) 
  \\
&- \frac{t_{\perp}^z}{3J_{\perp}/8}\chi^{z}_{\perp}  \sum_{i} \big(b_{z^21}^{\dagger}(i) b_{z^22}(i) +\text{h.c.}\big) 
-\chi^{xz}_{\parallel} \sum_{\langle i,j\rangle \alpha} 
\big( b_{x^2\alpha}^{\dagger}(i) b_{z^2\alpha}(j) 
+b_{z^2\alpha}^{\dagger}(i) b_{x^2\alpha}(j) 
+\text{h.c.}\big)  
\end{aligned}
\end{equation}
\end{widetext}
where the chemical potentials are given by, $\mu_b=\lambda_{z^2}+\lambda_{\text{tot}}$ for $3d_{z^2}$ orbital and $\mu_b+\delta\mu=\lambda_{x^2}+\lambda_{\text{tot}}$ for $3d_{x^2-y^2}$ orbital.
They fix the total density constraint. 

For the spinon part, the mean-field result is,
\begin{widetext}
\begin{equation}
\begin{aligned}\label{H_spinon}
&H_{\text{MF}}^{(f)}
=(\Delta_g -\delta\mu-\mu_f) \sum_{i\alpha\sigma} 
f_{x^2\alpha\sigma}^{\dagger}(i) f_{x^2i\alpha\sigma}(i) 
-(\delta\mu+ \mu_f) \sum_{i\alpha\sigma} 
f_{z^2\alpha\sigma}^{\dagger}(i)  f_{z^2\alpha\sigma}(i)  \\
&-\Big(\kappa^{x}_{\parallel} + \chi^{x}_{\parallel} \Big) \sum_{\langle i,j\rangle \alpha \sigma} \big( f_{x^2\alpha\sigma}^{\dagger}(i)  f_{x^2\alpha\sigma}(j)   
+\text{h.c.}\big) 
-\kappa^{z}_{\parallel} \sum_{\langle i,j\rangle \alpha \sigma} \big( f_{z^2\alpha\sigma}^{\dagger}(i) f_{z^2\alpha\sigma}(j)  +\text{h.c.}\big)
  \\
&- \kappa^{xz}_{\parallel} \sum_{i \alpha} 
\big( f_{x^2\alpha\sigma}^{\dagger}(i) f_{z^2\alpha\sigma}(i+x)  
+f_{z^2\alpha\sigma}^{\dagger}(i) f_{x^2\alpha\sigma}(i+x) 
 -f_{x^2\alpha\sigma}^{\dagger}(i) f_{z^2\alpha\sigma}(i+y)  -f_{z^2\alpha\sigma}^{\dagger}(i) f_{x^2\alpha\sigma}(i+y) 
+\text{h.c.}\big)  \\
&- \sum_{i\sigma} 
\big( \chi^{x}_{\perp} f_{x^21\sigma}^{\dagger}(i) f_{x^22\sigma}(i)
+ \chi^{z}_{\perp} f_{z^21\sigma}^{\dagger}(i) f_{z^22\sigma}(i)
+\text{h.c.}\big)
-\sum_{i} \Big[\Delta^{x}_{\perp}
\big(f_{x^21\uparrow}^{\dagger}(i) f_{x^22\downarrow}^{\dagger}(i)
-f_{x^21\downarrow}^{\dagger}(i) f_{x^22\uparrow}^{\dagger}(i) \big) 
+\text{h.c.}  \Big]	    \\
&-\sum_{i} \Big[\Delta^{z}_{\perp}
\big(f_{z^21\uparrow}^{\dagger}(i) f_{z^22\downarrow}^{\dagger}(i) 
-f_{z^21\downarrow}^{\dagger}(i) f_{z^22\uparrow}^{\dagger}(i) \big) 
+\text{h.c.}  \Big]
-
\sum_{\langle i,j\rangle\alpha} \Big[
\Delta^{x}_{\parallel}  \big(f_{x^2\alpha\uparrow}^{\dagger}(i) f_{x^2\alpha\downarrow}^{\dagger}(j) 
-f_{x^2\alpha\downarrow}^{\dagger}(i) f_{x^2\alpha\uparrow}^{\dagger}(j) \big) 
+\text{h.c.}
\Big]
\end{aligned}
\end{equation}
\end{widetext}
with the chemical potentials for the two orbitals $\mu_f=\lambda_{z^2}$, $\mu_f+\delta\mu=\lambda_{x^2}+\lambda_{\text{tot}}$.

\clearpage
\twocolumngrid
\bibliography{references}

\end{document}